\documentstyle[preprint,tighten,aps]{revtex}
\begin{document}
\draft
\title{Supersymmetry in quantum chaos and mesoscopic physics}
\author{Konstantin Efetov}
\address{Max Planck Institut f\"ur \\Festk\"orperforschung, Heisenbergst.1,
Stuttgart
70569, Germany and \\ L.D. Landau Institute for\\Theoretical Physics,
Moscow, Russia}
\date{\today{}}
\maketitle

\begin{abstract}
A brief review of the supersymmetry method and its application to mesoscopic
physics and quantum chaos is given. Alghough a non-linear supermatrix $%
\sigma $-model in this approach was derived from models with random
potential, it is emphasized that the zero-dimensional version of the $\sigma
$-model is equivalent to the random matrix theory and can even be derived
from the latter, too. This gives a possibility to use the zero -dimensional
model for description of problems of quantum chaos and mesoscopic physics. A
number of problems considered recently is presented. This includes nuclear
magnetic resonance in small metal particles and statistics of conductance
fluctuations in quantum dots. The solution of these problems became possible
due to a new possibility to calculate distribution functions.

\end{abstract}

\section{Introduction}

During last 10-15 years a significant progress has been achieved in
understanding quantum systems whose classical analogs exhibit chaotic
dynamics \cite{Gian,Gut}. Typical examples of such systems are quantum
billiards. In these models electrons are assumed to be confined in some
region of space. Depending on the shape of the billiard one can have both
the integrable and non-integrable models. In the latter ones the distance in
the phase space between any two trajectories even with very close initial
conditions diverges exponentially in time which is not the case in the
integrable models.

It turns out that quantum counterparts of the classically chaotic models
show completely different behavior from that of the models integrable in the
classical limit. In the chaotic models the energy levels are located in a
random way and it is quite reasonable to speak of level statistics. The
procedure of the averaging in such systems is the following. One fixes an
energy and studies the spectrum in an interval near this energy. Taking
other values of the energy one gets other parts of the spectrum. The
averaging procedure consists of averaging of some functions of the energy
levels calculated in the intervals near the fixed energies over these
energies.

A lot of numerical work has been carried out demonstrating that the energy
level statistics of the chaotic systems is well described by the random
matrix theory (RMT) proposed long ago \cite{Wig,Dy} in nuclear physics for
description of spectra of complex nuclei. One can find a complete account of
the RMT in the book of Mehta \cite{Meh}. In complex nuclei a large number of
particles interacts in an unknown way and therefore it is assumed that all
interactions are possible with some non-zero probability. According to the
basic statistical hypothesis of Wigner \cite{Wig} all matrix elements $%
H_{mn} $ of the Hamiltonian of the system having $N$ energy levels are
statistically independent and can take any values with an equal probability.
To make all integrals convergent it is necessary to cut them at large $%
H_{mn} $ corresponding to strong interactions and this should be done in an
invariant way. Wigner introduced the so-called Gaussian ensembles in which a
physical system having $N$ quantum states contributes with the following
statistical weight $D\left( H\right) $%
\begin{equation}
\label{in1}D\left( H\right) =A\exp \left[ -\sum_{m,n}^N\frac{\left|
H_{mn}\right| ^2}{a^2}\right]
\end{equation}
In Eq. (\ref{in1}) $a$ is the cutoff excluding strong interactions, $A\,$ is
the renormalization constant.

The symmetry of the matrix $H_{nm}$ depends on the symmetry of the system
under consideration. A system with the time reversal invariance and central
symmetry should be described by real symmetric matrices $H_{mn}$ and
corresponding ensemble is called Gaussian Orthogonal Ensemble (GOE). If the
time reversal symmetry is broken the matrices $H_{mn}$ should be Hermitian,
the corresponding ensemble being called Gaussian Unitary Ensemble (GUE).
Systems possessing the time reversal symmetry but lacking the central one
are described by matrices $H_{mn}$ with the elements being real quaternions.
The corresponding ensemble is called Gaussian Symplectic Ensemble (GSE).

Calculation of integrals over the $N\times N$ matrices, one should carry out
to get physical quantities, is not simple and it took quite a long time to
obtain the quantities of interest. By now a very well developed machinery
giving a possibility to make computation of the integrals exists \cite{Meh}
and many physical quantities have been calculated. As an example let us
write the results for the two-level correlation function $R\left( \omega
\right) $ defined as
\begin{equation}
\label{in2}R\left( \omega \right) =\left\langle
\mathop{\rm tr}
\delta \left( \epsilon -H\right)
\mathop{\rm tr}
\left( \epsilon -\omega -H\right) \right\rangle
\end{equation}
where the brackets $<...>$ stand for the averaging with the weight $D\left(
H\right) $ introduced in Eq. (\ref{in1}). The results are very well known
and can be written as
\begin{equation}
\label{in3}R_{orth}\left( x\right) =1-\frac{\sin ^2x}{x^2}-\frac d{dx}\left(
\frac{\sin x}x\right) \int_1^\infty \frac{\sin xt}tdt
\end{equation}
\begin{equation}
\label{in4}R_{unit}\left( x\right) =1-\frac{\sin ^2x}{x^2}
\end{equation}
\begin{equation}
\label{in5}R_{sympl}\left( x\right) =1-\frac{\sin ^2x}{x^2}+\frac d{dx}%
\left( \frac{\sin x}x\right) \int_0^1\frac{\sin xt}tdt
\end{equation}
where $x=\pi \omega /\Delta ,$ $\Delta $ is the mean energy level spacing.

Some time ago using a semiclassical approximation it was demonstrated also
analytically \cite{Ber,Smy,Arg} that the level-correlation functions for the
non-integrable models have the asymptotic behavior that agrees with the one
obtained from the RMT. Both the numerical and analytical study of the
quantum chaos have lead to the situation that now, although a general
analytical proof is absent, there is a believe that the level statistics of
quantum counterparts of the classically chaotic systems is well described by
the RMT.

Within condensed matter physics interest in quantum chaos was greatly
stimulated by recent advances in nanotechnology that made a fabrication of
very small and clean devices possible. In such experimentally available
systems electrons at low temperatures are scattered mainly by the boundaries
of the system and not by defects located inside (for a review see refs.\cite
{Bee,Alt}. These devices often called ''quantum dots'' resemble the quantum
billiards and therefore the theories of the quantum chaos are the first
candidates for their description \cite{Jal}. Computer modeling of the
transport through the quantum dots is quite popular and many interesting
results can be obtained in this way. As concerns analytical study two
approaches mentioned above, namely, semiclassical treatment and the RMT are
often used. Although the semiclassical approach is most direct this scheme
is rather complicated and is valid for calculation of correlations at
different energies at large energies only. As concerns the RMT it is a
purely phenomenological theory and it is not always clear how to express
physical quantities in terms of the random matrices. For example, these
difficulties arise if one wants to calculate correlations of currents,
dipole moments, etc. which cannot be directly expressed through the
matrices. Besides, one has to calculate difficult integrals over an
arbitrarily large number of variables. Therefore, often the problem cannot
be solved using the RMT. It is also not clear from such considerations what
happens it one adds impurities or changes the system in another way.

Fortunately, another method originally invented \cite{Efe} for studying
disordered metal exists which enables us to describe the quantum systems in
the regime of chaotic dynamics. It is based on the use of both commuting and
anticommuting Grassmann variables. Due to a symmetry between these variables
in the formalism developed it was called the supersymmetry method. In the
theory of disordered metals one starts with a Hamiltonian $H$ describing a
particle moving in an external potential $U\left( r\right) $ describing
impurities
\begin{equation}
\label{in6}H=H_0+U\left( r\right)
\end{equation}
where $H_0$ is the kinetic energy.

In order to calculate physical quantities one has to solve the Schr\"odinger
equation for an arbitrary potential $U\left( r\right) ,$ substitute the
eigenfunctions and eigenenergies into the corresponding formulae for the
physical quantities and, finally, to average over the random potential of
the impurities. In practice, this procedure can be carried out usually using
the perturbation theory in the limit of weak disorder only. As concerns,
phenomena like e.g. electron localization where such expansions are not
sufficient one encounters difficulties.

In the supersymmetry formalism one writes the physical quantities in such a
way that the averaging over the random potential can be done in the
beginning before all other manipulations. The price one pays for this
possibility is the use of the additional Grassmann variables. In fact, it
means only that one is to learn something more but manipulations with
quantities like supervectors or supermatrices containing both the commuting
and anticommuting variables are not more difficult than those with
conventional vectors and matrices. As a result of the averaging one obtains
a regular model without a randomness which is very similar to models of
statistical physics and field theory and, therefore, one can use very well
developed methods for further study. This model has a form of a non-linear $%
\sigma $-model and the free energy functional $F$ can be written as
\begin{equation}
\label{in7}F=\frac{\pi \nu }8\int
\mathop{\rm str}
\left[ D\left( \nabla Q\right) ^2+2i\omega \Lambda Q\right] dr
\end{equation}
In Eq. (\ref{in7}) $Q$ is an $8\times 8$ supermatrix, a matrix containing in
half both commuting and anticommuting elements, with the constraint $Q^2=1.$
The parameter $\nu $ in Eq. (\ref{in7}) is the density of states which is
assumed to be a constant, $D$ is the classical diffusion coefficient and $%
\omega $ is the frequency. The symbol $%
\mathop{\rm STr}
$ stands for a supertrace which is an analog of the trace for the
conventional matrices and $\Lambda $ is the $8\times 8$ supermatrix of the
form
\begin{equation}
\label{in8}\Lambda =\left(
\begin{array}{cc}
1 & 0 \\
0 & -1
\end{array}
\right)
\end{equation}
For studying transport one has to calculate usually a functional integral $I$
which can be written in a symbolic form as
\begin{equation}
\label{in9}I=\int QQ\exp \left( F\left[ Q\right] \right) DQ
\end{equation}
In Eq. (\ref{in9}) the symbolic product $QQ$ means that products of two
matrix elements of the supermatrices $Q$ are taken. These elements can be at
different points. The symbolic form Eq. (\ref{in9}) is used to avoid
specifying different details which are not necessary now. Depending on the
symmetries of the physical system the supermatrix $Q$ has three different
types of the symmetry that correspond to three different ensembles in the
RMT.

The free energy functional $F$ Eq. (\ref{in7}) is very similar to the
classical Heisenberg model, the frequency $\omega $ playing the role of an
external magnetic field. The only difference is that now one deals with
supermatrices instead of vectors. Depending on the dimensionality $d$ of the
space in Eq. (\ref{in7}) one can study such phenomena as the Anderson
metal-insulator transition for $d>2$ or localization in films ($d=2)$ and
wires ($d=1).$ These problems are not discussed here because they are out of
scope of the conference. As concerns mesoscopic physics one should use the
zero dimensional $\sigma $-model. This can be understood very easily. In a
limited volume one can expand the supermatrix $Q$ as a function of
coordinates in a series in spatial harmonics. The set of the eigenenergies
corresponding to these harmonics is discrete and provided the inequality
\begin{equation}
\label{in10}\omega \ll E_c=\pi ^2D/L^2
\end{equation}
is fulfilled one can retain the zero harmonics only. Then the free energy
functional $F\left[ Q\right] $ Eq. (\ref{in7}) is substituted by the
function $F_0\left( Q\right) $%
\begin{equation}
\label{in11}F_0\left( Q\right) =\frac{i\pi \omega }{4\Delta }%
\mathop{\rm STr}
\left( \Lambda Q\right)
\end{equation}
where $\Delta =\left( \nu V\right) ^{-1}$ is the mean level spacing. Now,
the supermatrix $Q$ does not depend on the space coordinate and one has in
Eq. (\ref{in9}) instead of the functional integral a definite one.
Calculation of the definite integral over several variables (usually a
reduction to integration over 2 or 3 variables depending on the ensemble
considered) is not a very difficult task and can be carried out analytically
in many interesting cases. As the first application of the supersymmetry
formalism the two-level correlation function $R$ has been calculated \cite
{Efe} and the result coincided with Eqs. (\ref{in3}-\ref{in5}). This was the
first analytical confirmation of the RMT starting from a microscopic model.

One can naturally ask why a theory constructed for a description of
disordered metals can be applied to quantum chaos and, in particular, to the
quantum billiards where the electron scattering is due to scattering by the
boundaries. Of course, a system with a complicated potential of impurities
belongs to the class of the system with classically chaotic behavior and,
provided the averaging over different configurations of the potential is
equivalent to the averaging over different parts of the spectrum, the
non-linear $\sigma $-model is appropriate at least for these chaotic models.
However, the question about the applicability of the $\sigma $-model for a
description of such non-integrable models as, for example, Sinai billiard is
more difficult. By now, any derivation of Eq. (\ref{in11}) for such models
starting from the first principles is not available.

Fortunately, one can derive the zero-dimensional $\sigma $- model (only
zero-dimensional!) from the Gaussian Ensembles of the RMT. Now, instead of
the averaging over the random potential of impurities one has to average
over the matrix elements of the Hamiltonian with the weight determined by
Eq. (\ref{in1}). One can find the derivation in Refs. \cite{Ver,Altl} which
is very similar to the one presented for the disordered metals \cite{Efe}
and can be carried out rigorously in the limit $N\gg 1.$ This result
establishes the equivalence of the RMT and the zero-dimensional $\sigma $%
-model. Thus, if one believes that the RMT describes well the quantum chaos
problems one may apply also the supersymmetry technique.

The calculations with the $\sigma $-model are not only more straightforward
than those within the RMT but can also be generalized to higher dimensions
where the RMT is not applicable. For example, one can consider long wires
and use for the description the one-dimensional $\sigma $-model. Recently,
it was found that many phenomena known in the field of the quantum chaos as
''dynamical localization'' \cite{Izr,Cas} are properly described by the
ensembles of Random Banded Matrices (RBM) \cite{CMI}. The model of (RBM) was
in its turn mapped onto the one-dimensional supermatrix $\sigma $-model \cite
{Fyo} which made possible to treat analytically many difficult problems
related to the dynamical localization.

An interesting new development in study of weakly disordered metallic
systems and their chaotic counterparts was done in Refs. \cite{Sim,Lee}. The
authors of this works studied how the energy levels change with changing an
external parameter. The role of such a parameter can be played by an
external magnetic field, a shape of the potential confining the quantum
billiard and any other physical parameter the Hamiltonian of the system
depends on. The calculation of the corresponding quantities was done with
the a slight modification of the supersymmetric $\sigma $-model. Due the
highly universal form of the zero-dimensional version of the model the final
formulae expressing the dependence of the energy levels on the parameter are
universal and this leads to a new class of universality describing systems
with the quantum chaotic behavior.

The number of works studying quantum chaos in mesoscopic systems with the
supersymmetry technique is fast growing because in many cases it is the only
way of analytical calculations and it is difficult to give even a brief
account of all these works here. Therefore, below only some directions of
the research I have personally been involved in last years are presented.
These works originated from a discovered possibility to calculate different
distribution functions characterizing mesoscopic fluctuations and have been
performed in cooperation with Vladimir Prigodin, Shinji Iida and Vladimir
Falko.

\section{Nuclear magnetic resonance in small metal particles}

In this Section I want to demonstrate that the chaotic electron motion in
small mesoscopic metal particles results in very interesting properties of
the nuclear magnetic resonance (NMR) provided the inelastic mean free path $%
l_\phi $ exceeds the sample size. It is assumed that in the corresponding
bulk metal the NMR line shape is very narrow and all effects considered are
due to the chaotic electron motion in the particles which can be due to the
scattering of the electrons either by impurities inside the particles or by
the boundaries. The main result is that the NMR line shape becomes very
broad and asymmetric when decreasing temperature or the particle size \cite
{EP}

Usually, when studying NMR in a bulk metal, it assumed that the local
paramagnetic susceptibility $\chi \left( r\right) $ determining the Knight
shift does not fluctuate considerably and therefore one can substitute this
quantity by the Pauly susceptibility. Such a substitution is correct
provided the mean free path $l$ is large, $lp_0\gg 1,$ where $p_0$ is the
Fermi momentum. However, a small particle size can lead to strong
fluctuations of the wave function. Due to the chaotic character of motion,
these fluctuations can be very strong even in clean particles where
electrons can move without collisions from boundary to boundary. The local
paramagnetic susceptibility $\chi \left( r\right) $ determining the local
Knight shift strongly depends on the wave function $\psi $ at the point $r$
and also fluctuates. One can start with a conventional formula for the shift
$\Delta \omega \left( r_a\right) $ of the resonance frequency of a nuclear
spin located at a point $r_a$%
\begin{equation}
\label{nu1}\Delta \omega \left( r_a\right) =J\left( g\mu \right) ^{-1}\chi
\left( r_a\right) H
\end{equation}
It is assumed in Eq. (\ref{nu1}) that the shift $\Delta \omega \left(
r_a\right) $ is due to the Fermi contact interaction of the nuclear spin
with spins of conduction electrons, $J$ being the hyperfine coupling
constant, $g\mu $ being the coefficient determining the Zeeman splitting of
the electron states. The function $\chi \left( r_a\right) $ is Eq. (\ref{nu1}%
) is the local spin susceptibility of the conduction electrons. In the
absence of electron-electron interactions the susceptibility $\chi \left(
r_a\right) $ can be expressed through the one-particle local density of
states $\rho \left( \epsilon ,r_a\right) $ in a standard way
\begin{equation}
\label{nu2}\chi \left( r_a\right) =2\left( g\mu \right) ^2\int \rho \left(
\epsilon ,r_a\right) f\left( \epsilon \right) d\epsilon
\end{equation}
where $f\left( \epsilon \right) =\left( 4T\right) ^{-1}\cosh ^{-2}\left(
\left( \epsilon -\epsilon _0\right) /2T\right) $ is the derivative of the
Fermi function, $\epsilon _0$ is the Fermi energy. At $T=0$ the function $%
f\left( \epsilon \right) $ in Eq. (\ref{nu2}) is just the $\delta $-function
and the local spin susceptibility $\chi \left( r_a\right) $ is proportional
to the local density of states $\rho \left( \epsilon ,r_a\right) .$

In a traditional bulk metal the shift $\Delta \omega $ is the same for all
nuclear spins and the NMR line is narrow. The local density of states $\rho
\left( \epsilon ,r_a\right) =\nu $ does not depend on coordinate and is
equal to
\begin{equation}
\label{nu3}\nu =mp_0/2\pi ^2
\end{equation}
where $p_0$ is the Fermi momentum. The local susceptibility $\chi \left(
r_a\right) $ coincides with the Pauli susceptibility $\chi _p$%
\begin{equation}
\label{nu4}\chi _p=2\left( g\mu \right) ^2\nu
\end{equation}
In disordered metals in the limit $lp_0\gg 1,$ where $l$ is the mean free
path, the local density of states and the susceptibility are again given by
Eqs. (\ref{nu3}, \ref{nu4}). Therefore, in this approximation, the NMR line
of the bulk metal is infinitely narrow. Now, let us try to answer the
question what happens if the bulk metal is substituted by a system of small
particles of the same metal and of equal size.

The starting formula for the resonance line shape $I\left( \omega \right) $
corresponding to a system of nuclear spins is written as
\begin{equation}
\label{nu5}I\left( \omega \right) =A\sum_a\delta \left( \omega -\omega
_0-\Delta \omega \left( r_a\right) \right)
\end{equation}
where the constant $A$ stands for a weight and $\omega _0$ is the
non-shifted position of the resonance. In Eq. (\ref{nu5}) one should sum
over positions $r_a$ of all nuclear spins in the system. We assume that all
the nuclear spins are located in a system of macroscopically equal
particles. At the same time it is assumed the each particle has its own
chaotic electron motion.

Comparing Eqs. (\ref{nu1}, \ref{nu2},\ref{nu5}) one can see that the NMR\
line shape coincides exactly with the local density of states distribution
function $P\left( x\right) $
\begin{equation}
\label{nu6}P\left( x\right) =\left\langle \delta \left( x-\rho \left(
\epsilon ,r\right) /\nu \right) \right\rangle
\end{equation}
where $<...>$ stands for impurities in the system.

The local density of states $\rho \left( \epsilon ,r\right) $ entering Eq. (%
\ref{nu6}) can be written in terms of the retarded $G^R$ and advanced $G^A$
Green functions as
\begin{equation}
\label{nu7}\rho \left( \epsilon ,r\right) =-\left( 2\pi i\right) ^{-1}\left(
G_\epsilon ^R\left( r,r\right) -G_\epsilon ^A\left( r,r\right) \right)
\end{equation}
Neglecting electron-electron interactions the Green functions can be written
in the form
\begin{equation}
\label{nu8}G_\epsilon ^{R,A}\left( r,r^{^{\prime }}\right) =\sum_\beta \frac{%
\psi _\beta \left( r\right) \psi _\beta ^{*}\left( r^{^{\prime }}\right) }{%
\epsilon -\epsilon _\beta \pm i\gamma /2}
\end{equation}
where $\psi _\beta \left( r\right) $ and $\epsilon _\beta $ are the
eigenfunctions and eigenenergies of an isolated metal particle. However, the
metal particles are assumed to be not completely isolated and there is a
finite probability for electrons to leave the particle. Such a situation can
correspond to a system of metal particles in a non-metallic matrix. Due to
the connection with the surrounding, the energy levels in the particle are
no longer discrete. This can correspond to the experimental systems
considered in Refs. \cite{Hal,Brom}. We describe the level width by the
parameter $\gamma $ entering Eq. (\ref{nu8}), which does not depend on $%
\beta .$ Due to the connection with the surrounding we may assume that the
chemical potential in the particles is fixed and does not vary when changing
the impurity configuration.

The possibility to write the local density of states distribution function $%
P\left( x\right) $ in terms of the Green functions enables us to use the
supersymmetry technique. This distribution function can be written in terms
of an integral over the supermatrix $Q$
\begin{equation}
\label{nu9}P\left( x\right) =\int_0^{2\pi }\frac{dt}{2\pi }\left\langle
\delta \left( x-\frac 12\left(
Q_{33}^{11}-Q_{33}^{22}+Q_{33}^{12}e^{-it}-Q_{33}^{21}e^{it}\right) \right)
\right\rangle _Q
\end{equation}
where the symbol $<...>_Q$ stands for the integral over the supermatrix $Q$
with the weight $\exp \left( -F_0\left[ Q\right] \right) $ and $F_0\left[
Q\right] $ is determined by the following equation
\begin{equation}
\label{nu9a}F_0\left[ Q\right] =-\left( \alpha /4\right)
\mathop{\rm STr}
\left( \Lambda Q\right)
\end{equation}
where $\alpha =\gamma \pi /\Delta $ and $Q_{mn}^{kl}$ denote elements of the
supermatrix $Q.$

Comparing Eq. (\ref{nu9a}) with Eq. (\ref{in11}) we see that now, instead of
the frequency $\omega $ the level width $\gamma $ enters the expression for $%
F_0\left[ Q\right] .$

Calculation of the integral in Eq. (\ref{nu9}) is rather straightforward and
the result for the unitary ensemble has the form
\begin{equation}
\label{nu10}P\left( x\right) =\left( \alpha /8\pi \right)
^{1/2}x^{-3/2}\left( 2\cosh \alpha +\left( x+1/x-1/\alpha \right) \sinh
\alpha \right) \exp \left[ -\alpha \left( x+1/x\right) /2\right]
\end{equation}

Eq. (\ref{nu10}) is the complete solution of the problem of calculation of
the local density of states distribution function. At small $\alpha $
corresponding to low temperatures or small particle sizes the function $%
P\left( x\right) $ is very broad and asymmetric. It is represented for
several values of $\alpha $ in Refs. \cite{EP}. Recently Eq. (\ref{nu10})
was obtained \cite{Bee2} using the calculational scheme of RMT \cite{Meh}.

The function $P\left( x\right) $ at $\alpha \rightarrow \infty $ approaches $%
P_0\left( x\right) =\delta \left( x-1\right) $ corresponding to the
distribution function of the bulk metal. In the limit of a small particle $%
\alpha \ll 1$ it has the following asymptotics
\begin{equation}
\label{nu11}P\left( x\right) =\left( \alpha ^3/8\pi \right)
^{1/2}x^{-5/2}\exp \left( -\alpha /2x\right) ,\;x\ll \alpha
\end{equation}
\begin{equation}
\label{nu12}P\left( x\right) =\left( \alpha /8\pi \right)
^{1/2}x^{-3/2},\;\alpha \ll x\ll \alpha ^{-1}
\end{equation}
\begin{equation}
\label{nu13}P\left( x\right) =\left( \alpha ^3/8\pi \right)
^{1/2}x^{-1/2}\exp \left( -\alpha x/2\right) ,\;x\gg \alpha ^{-1}
\end{equation}
These asymptotics can be reproduced using simple assumptions about
fluctuations of the electron wave functions which gives a possibility to get
an information about the fluctuations. It turns out that Eqs. (\ref{nu11},
\ref{nu13}) can be obtained (for a detailed derivation see the second Ref.
\cite{EP}) assuming that the probability for a wave function $\psi _\beta
\left( r\right) $ at a point $r$ to have the square of the modulus $|\psi
|_\beta ^2$ is proportional to
\begin{equation}
\label{nu14}\exp \left( -V|\psi |_\beta ^2\right)
\end{equation}
where $V$ is the volume.

Eq. (\ref{nu12}) can also be understood in simple terms and the
corresponding discussion is presented in Ref. \cite{EP}.

Repeating the same steps of the calculations one can obtain all moments $%
M_n\left( \alpha \right) $ of the local density of states (of course, they
can be obtained from the distribution function, too). The result reads
\begin{equation}
\label{nu15}M_n\left( \alpha \right) =\frac n{\left( 2\alpha \right) ^{n-1}}%
\frac{\left( 2n-2\right) !}{\left( n-1\right) !}
\end{equation}
Using Eq. (\ref{nu15}) one can derive the coefficients $b_n$ of the
participation ratio defined as
\begin{equation}
\label{nu16}b_n=\nu ^{-n}\left\langle \sum_\beta |\psi _\beta \left(
r\right) |^{2n}\delta \left( \epsilon -\epsilon _\beta \right) \right\rangle
\end{equation}
The result is rather simple
\begin{equation}
\label{nu17}b_n=\Delta ^{n-1}n!
\end{equation}
and agrees with the assumed form Eq. (\ref{nu14}) of the distribution of the
wave functions.

Non-zero temperature effects can be incorporated into the theory presented
above. It turns out that the maximum of the NMR line linearly changes with
changing the temperature, the result claimed to have been observed long ago
\cite{Yee}. Some other characteristic features of the effects considered
like e.g. asymmetry of the line shape have also been observed in old
experiments (see review \cite{Hal}). Quite recently \cite{Brom2} the
relevance of the theory to Pt clusters was demonstrated in Leiden. The
discussion presented shows that the NMR can be an interesting possibility to
study mesoscopic effects and there is no doubt that both the theoretical and
experimental investigations will continue.

\section{Statistics of conductance fluctuations in quantum dots.}

Study of transport through very small systems in which quantum effects are
extremely important is another type of experiments that attract a lot of
interest now \cite{Kirk}. The tunneling spectroscopy in these novel objects
is done by measuring conductance versus gate voltage, magnetic field or
other external parameters. The most striking feature of the experiments on
the mesoscopic structures is an irregular dependence of of the conductance
on the varied parameters \cite{Kas,Mar}. The conductance fluctuations are
typically of the same order of magnitude as the average conductance.
Therefore, in order to describe an experiment in an adequate way one needs
to know not only traditional averages but the whole distribution function.

The quantum dots can be modeled by quantum billiards and a statistical
theory of conductance fluctuations for non interacting electrons can be
constructed using again the supersymmetry method \cite{Pri}. One can start
from a model of a dot with two point contacts. Its effective Hamiltonian can
be written in the form
\begin{equation}
\label{st1}H=H_c\pm \frac i{2\pi \nu }\left[ \alpha \nu \Delta +\alpha
_1\delta \left( r-r_1\right) +\alpha _2\delta \left( r-r_2\right) \right]
\end{equation}
where $H_c$ stands for the Hamiltonian of the closed dot
\begin{equation}
\label{st2}H_c=H_0+H_1
\end{equation}
with $H_0$ being the kinetic energy and $H_1$ being a confinement potential
which contains an irregular part due to random impurities and imperfections
of the dot shape. In Eq. (\ref{st1}) $\nu $ is the mean density of states
and $\Delta =\left( \nu V\right) ^{-1}$ is the mean level spacing, with $V$
being the volume.

The last term in Eq. (\ref{st1}) stands for the contacts and can be derived
starting with the term $\left[ J\psi _l^{+}\left( r\right) \psi \left(
r\right) +h.c.\right] $ describing tunneling between the dot and the
external leads (see also a more general derivation in Refs. \cite{Iid,Zir}).
Approximating by the point contacts is valid if the size of the contact is
of the order of one electron wave. The terms describing three contacts are
written in Eq. (\ref{st1}), the first term being correspondent to an
additional extended lead, and the last two to the point contacts the
electrons tunnel through. One should take the sign $+$ in Eq. (\ref{st1})
when calculating the advanced Green function $G^A$ and the sign $-$ for the
retarded one $G^R.$

Explicit calculations can be carried out for an arbitrary relation between $%
\alpha _1,\alpha _2$ and $\alpha .$ However, for the sake of simplicity let
us consider the case $\alpha \gg \alpha _{1,2}.$ In this approximation the
level width $\gamma =\alpha \Delta /\pi $ is the same for all levels.

Although the level width $\gamma $ does not fluctuate there can be strong
conductance fluctuations due to fluctuations of wave functions and
eigenenergies. One can write the conductance $G$ using a Landauer type
formula
\begin{equation}
\label{st3}G=\frac{2e^2}h\frac{\alpha _1\alpha _2}{\left( \pi \nu \right) ^2}%
G_\epsilon ^R\left( r_1,r_2\right) G_\epsilon ^A\left( r_2,r_1\right)
\end{equation}
The distribution function $B\left( g\right) $ of the conductances is defined
as
\begin{equation}
\label{st4}B\left( g\right) =\left\langle \delta \left( g-Gh/2e^2\right)
\right\rangle
\end{equation}
where angular brackets stand for averaging over irregularities in the
system. After some, by now standard for the supersymmetry method,
transformations Eq. (\ref{st4}) can be reduced to the form
\begin{equation}
\label{st5}B\left( g\right) =\left\langle \delta \left( g+\alpha _1\alpha
_2Q_{33}^{12}\left( r_1\right) Q_{33}^{21}\left( r_2\right) \right)
\right\rangle _Q
\end{equation}
Again, as in the preceding section, the symbol $<...>_Q$ means integration
with $\exp \left( -F_0\left[ Q\right] \right) ,$ $F_0\left[ Q\right] $ being
determined by Eq. (\ref{nu9a}). Calculating the integral in Eq. (\ref{st5})
for the unitary ensemble one obtains the final result
\begin{equation}
\label{st6}B\left( g\right) =-\frac 1{2\alpha \bar g\lambda }\frac d{%
d\lambda }\left[ \frac 2\lambda e^{-\alpha \lambda }\cosh \alpha +\left( 1-%
\frac 1{\alpha \lambda }+\frac 1{\lambda ^2}+\frac 1{\alpha \lambda ^3}%
\right) e^{-\alpha \lambda }\sinh \alpha \right]
\end{equation}
where
\begin{equation}
\label{st7}\lambda =\sqrt{1+2g/\left( \alpha \bar g\right) },\;\bar g%
=2\alpha _1\alpha _2/\alpha
\end{equation}
Eqs. (\ref{st6}, \ref{st7}) describe completely the statistics of
conductance fluctuations in the quantum dot. Any averages can be calculated
using the distribution function $B\left( g\right) .$ For example, one can
check easily that $\bar g$ in Eq. (\ref{st7}) is just the average
conductance of the dot. The function $B\left( g\right) $ is represented for
several values of $\alpha $ in Ref. \cite{Pri}.

Surprizingly it decreases monotonously and $g=0$ is the most probable value
of the conductance. Such a behavior is very different from that known for
samples strongly connected with leads \cite{ALS}, where the maximum of the
distribution function corresponds to a finite value of the conductance. It
is also different from the behavior of the distribution function of the
density of states, considered in the preceding section, that approaches the $%
\delta $-function already in the limit $1\ll \alpha \ll E_c/\Delta ,$
whereas $B\left( g\right) $ Eqs. (\ref{st6}, \ref{st7}) in this limit takes
for $g\ll \alpha \bar g$ the form
\begin{equation}
\label{st9}B\left( g\right) =\left( 1/g\right) \exp \left( -g/\bar g\right)
\end{equation}
The exponential behavior Eq. (\ref{st9}) gives a very interesting
information about the phases $\phi _\beta \left( r_i\right) $ of the wave
functions of the closed dot, namely the phases $\phi _\beta =\phi _\beta
\left( r_1\right) -\phi _\beta \left( r_2\right) $ for different states $%
\beta $ are completely uncorrelated \cite{Pri}.

In the limit $\alpha \ll 1$ the asymptotics of the function $B\left(
g\right) $ can be written in the form
\begin{equation}
\label{st10}B\left( g\right) =\frac{\alpha ^2}{2\bar g}\left\{
\begin{array}{cc}
4/\alpha ^3, & g\ll \alpha
\bar g, \\ u^{-3/2}, & \alpha
\bar g\ll g\ll \bar g/\alpha \\ u^{-1/2}\exp \left( -u^{1/2}\right) , & \bar
g/\alpha \ll g,
\end{array}
\right.
\end{equation}
where $u=2\alpha g/\bar g.$

The non-zero probability for a finite $g$ in the region $g\gg \bar g/\alpha $
in Eq. (\ref{st10}) signals as in the preceding Section about very strong
fluctuations of the wave functions $\psi _\beta $ and can be reproduced in
the same way as the corresponding asymptotics of the distribution function
of the density of states. It turns out that the assumption Eq. (\ref{nu14})
about the Gaussian distribution of the amplitude of the wave function is
sufficient to understand the exponential decay in Eq. (\ref{st10}).

Knowing the statistical properties of the wave function one can describe all
the transport properties of the quantum dot. Depending on the quality of the
contacts one can distinguish between two class of experiments on quantum
dots. If the connection to the dot is not very good, electron tunneling into
the dot is a rare event. Adding one electron into the dot causes an increase
of the electrostatic energy by $e^2/C,$ where $C$ is the capacitance of the
dot, which is a large energy. It drastically diminishes the tunneling
probability unless a special gate voltage
\begin{equation}
\label{st11}V_N=\left( e/C\right) \left( N+1/2\right)
\end{equation}
is applied \cite{Kas}. This is a so called Coulomb blockade effect. In dots
with good contacts \cite{Mar} the probability of tunneling through the dot
is larger and the Coulomb blockade effects seem to be not as important. The
latter class of the experiments can be described by the theory of
non-interacting electrons developed above. At first glance, for a
description of the former class of the experiments one needs a theory of
interacting electrons. However, as it was noticed in Ref. \cite{JBS} the
only thing one needs to describe the most interesting feature of the
experiments, namely, the fluctuations of the conductance peaks, is the
distribution function of the wave functions. This is due to the fact that
the conductance peaks correspond to the resonance tunneling through one
level in the dot at the gate voltages determined by Eq. (\ref{st11}). For a
dot with two point contacts the resonance conductance can be written as
\begin{equation}
\label{st12}g_m=\int \frac{d\epsilon }{4T}\frac{\gamma _1\gamma _2}{\epsilon
^2+\left( \gamma _1+\gamma _2\right) ^2/4}\cosh ^{-2}\frac \epsilon {4T}
\end{equation}
where $\gamma _i=\alpha _iV|\psi _\beta \left( r_i\right) |^2\Delta /\pi ,$ $%
V$ being the volume of the dot, $T$ is the temperature, the subscript $\beta
$ corresponds to the resonance level. Knowing the distribution function of
the wave functions one can make an appropriate averaging and obtain a
distribution function $R\left( g_m\right) $of the resonance conductances $%
g_m.$ Using Eq. (\ref{nu14}) a general result in the limit $\alpha
_{1,2}\Delta \ll T\ll \Delta $ is reduced to the form \cite{Pri}
\begin{equation}
\label{st13}R\left( g_m\right) =x^2\exp \left( -ax\right) \left[ K_0\left(
x\right) +aK_1\left( x\right) \right] /g_m,
\end{equation}
where
\begin{equation}
\label{st14}x=\left( 4T/\Delta \right) \left( g_m/\sqrt{\alpha _1\alpha _2}%
\right) ,\qquad a=1+\frac 12\left[ \left( \alpha _1/\alpha _2\right)
^{1/4}-\left( \alpha _2/\alpha _1\right) ^{1/4}\right] ^2
\end{equation}
and $K_{0,1}$ are the Bessel functions.

The distribution function $R\left( g_m\right) $ can also be calculated in
the opposite limit $T\ll \alpha _{1,2}\Delta $ and reads as
\begin{equation}
\label{st15}R\left( g_m\right) =\frac 1{2\sqrt{1-g_m}}\frac{1+\left(
2-g_m\right) \left( a^2-1\right) }{\left[ 1+g_m\left( a^2-1\right) \right] ^2%
}
\end{equation}
The results of the present Section show that the supersymmetry method can be
very useful for describing transport in quantum dots. Essential quantities
that can be directly compared with experiments are obtained in a regular
way. In the discussion presented above Eq. (\ref{nu14}) was used several
times. It was obtained for the unitary ensemble from studying asymptotics of
the distribution functions of the density of states or conductances. The
distribution function of the wave functions is an important quantity and in
the next Section it will be derived directly for an arbitrary magnetic
field, thus providing a description of the crossover between the unitary and
orthogonal ensemble including the limit of the zero field.

\section{Statistics of fluctuations of wave functions in arbitrary magnetic
field.}

The distribution function of the amplitude of the wave functions Eq. (\ref
{nu14}) was obtained in the previous Sections in indirect ways calculating
the density of states and conductance distribution functions, then,
calculating all moments, reconstructing the coefficients of the
participation ratio and, on their basis, the whole distribution function.
The form of the distribution function in the same as the one known in the
RMT for the unitary ensemble \cite{Bro}. The other limiting case of the
orthogonal ensemble is also known in the RMT whereas the crossover between
the ensembles has never been studied. It is quite natural to try to check
the prediction of the RMT for the orthogonal ensemble and describe the
crossover using the supersymmetry method. The indirect way of calculations
used in the preceding Sections is not very convenient and now a more
straightforward method \cite{Fal} will be used.

The distribution function $f\left( t\right) $ of the amplitude of wave
functions $\psi _\beta \left( r\right) $ normalized to the volume $V$ is
defined as
\begin{equation}
\label{wf1}f\left( t\right) =\Delta \left\langle \sum_\beta \delta \left(
t-|\psi _\beta \left( r\right) |^2V\right) \delta \left( \epsilon -\epsilon
_\beta \right) \right\rangle
\end{equation}
where, as previously, the angle brackets denote averaging over
irregularities of the system and the sum is extended over all states in the
quantum box. Calculating the moments of the distribution function $f\left(
t\right) $ one can obtain the participation ratio coefficients. In order to
use the supersymmetry method one has to express, first, the distribution
function in terms of the retarded and advanced Green functions Eq. (\ref{nu8}%
). It can be done expanding the delta-function in Eq. (\ref{wf1}) in a
series in the fluctuating quantity $\tilde t=V|\psi _\beta \left( r\right)
|^2.$ Using the discreteness of the spectrum of the quantum finite system
and Eq. (\ref{nu8}) we can relate each participation ratio $\left( \tilde t%
\right) ^n$ to the expression
$$
\left( i\gamma V\right) ^n\int_{-E/2}^{E/2}\frac{d\epsilon }{\nu E}\int
\frac{dr^{^{\prime }}}{2\pi i}\left( G_\epsilon ^A\left( r^{^{\prime
}},r^{^{\prime }}\right) -G_\epsilon ^R\left( r^{^{\prime }},r^{^{\prime
}}\right) \right) \left[ G_\epsilon ^R\left( r,r\right) \right] ^n
$$
taken in the limit when the energy level smearing $\gamma $ is much smaller
than the mean level spacing $\Delta .$ After that one can recollect the
series back and arrive at
\begin{equation}
\label{wf2}f\left( t\right) =\Delta \lim _{\gamma \rightarrow 0}\left\langle
\int \frac{dr^{^{\prime }}}{2\pi i}\left( G_\epsilon ^A\left( r^{^{\prime
}},r^{^{\prime }}\right) -G_\epsilon ^R\left( r^{^{\prime }},r^{^{\prime
}}\right) \delta \left( t-i\gamma VG_\epsilon ^R\left( r,r\right) \right)
\right) \right\rangle
\end{equation}
At first glance, the argument of the $\delta $-function that appears in Eq. (%
\ref{wf2}) is complex at finite $\gamma $ which would imply that this
function is just a formal series. However, in the limit $\gamma \rightarrow
0,$ this argument becomes real as can be seen from the form of the Green
function given by Eq. (\ref{nu8}). Using Eq. (\ref{wf2}) one can already
apply the supersymmetry technique and after standard manipulations we obtain
\begin{equation}
\label{wf3}f\left( t\right) =\frac 18\int_0^{2\pi }\int_0^{2\pi }\frac{%
d\zeta _1d\zeta _2}{\left( 2\pi \right) ^2}\lim _{\gamma /\Delta \rightarrow
0}\left\langle \left( \bar v_1Qv_1\right) \delta \left( t-\frac{\gamma
\left( \bar v_2Qv_2\right) }{4\Delta }\right) \right\rangle _Q
\end{equation}
where the $8$-component vectors $\bar v_1$ and $\bar v_2$ are: $\bar v_1=%
\sqrt{2}\left( 0,0,e^{i\zeta _1},e^{-i\zeta _1},0,0,0,0\right) ,$ $\bar v_2=%
\sqrt{2}\left( 0,0,0,0,0,0,e^{i\zeta _2},e^{-i\zeta _2}\right) $ and by the
angular brackets integration over $Q$ with the weight $\exp \left(
-F_0\left[ Q\right] \right) $ is denoted. In the limiting case of high
magnetic fields one should integrate over the matrices $Q$ corresponding to
the unitary ensemble and take $F_0\left[ Q\right] $ from Eq. (\ref{nu9a}).
The case of an arbitrary field needs some modifications. The general
situation can be described by integration over the matrices $Q$ with the
symmetry corresponding to the orthogonal ensemble but now, instead of Eq. (%
\ref{nu9a}) the following form of the free energy $\tilde F_0\left[ Q\right]
$ should be used
\begin{equation}
\label{wf4}\tilde F_0\left[ Q\right] =F_\phi \left[ Q\right] +F_0\left[
Q\right] ,
\end{equation}
$$
F_\phi \left[ Q\right] =-\left( X/4\right) ^2%
\mathop{\rm Str}
\left[ \left( \left[ Q,\tau _3\right] \right) ^2\right]
$$
In Eq. (\ref{wf4}) $\left[ Q,\tau _3\right] $ is a commutator of $Q$ with a
matrix $\tau _3$ serving to violate a symmetry of the supermatrix $Q$
corresponding to the time reversal symmetry, the parameter $X$ is
proportional to the magnetic field applied
\begin{equation}
\label{wf5}X^2=\frac{2\pi D}\Delta \int \frac{d{\bf r}}V\left( \frac{2\pi
{\bf A}\left( {\bf r}\right) }{\phi _0}\right) ^2=\alpha _g\frac{\phi ^2}{%
\phi _0^2}\frac{E_c}\Delta
\end{equation}
where ${\bf A}$ is the vector potential, $\phi _0$ is the flux quantum and $%
\alpha _g$ is a numerical factor depending on the geometry, $D$ is the
diffusion coefficient and $E_c$ is the so called Thouless energy introduced
in Eq. (\ref{in10}). Calculation of the integral over $Q$ in Eq. (\ref{wf3})
can been done using the parametrization described in Ref. \cite{Altl} and
after a rather lengthy algebra one comes to the final form of the
distribution function $f\left( t\right) $
\begin{equation}
\label{wf6}f\left( t\right) =2\int_1^\infty \left[ \Phi _1\left( X\right)
\left( \left( xX\right) ^2-1\right) +\Phi _2\left( X\right) \right] \left(
xX\right) ^2I_0\left( tx\sqrt{x^2-1}\right)
\end{equation}
$$
\times \exp \left( -tx^2-X^2\left( x^2-1\right) \right) dx
$$
where
\begin{equation}
\label{wf6a}\Phi _1\left( X\right) =\frac{e^{-X^2}}X\int_0^Xe^{y^2}dy,\qquad
\Phi _2\left( X\right) =\frac{1-\Phi _1\left( X\right) }{X^2}
\end{equation}

In the above expression $I_0\left( z\right) $ is the modified Bessel
function and all the magnetic field dependence is included in the parameter $%
X.$

Using the general expression Eq. (\ref{wf6}) for the distribution function $%
f\left( t\right) $ one can obtain the asymptotics in the limits $%
X\rightarrow 0$ and $X\rightarrow \infty $ which correspond to the
orthogonal and unitary ensembles. In this limits the integration over $x$ in
Eq. (\ref{wf6}) can easily be done and one can obtain
\begin{equation}
\label{wf7}f_{X\rightarrow 0}\left( t\right) =\frac{e^{-t/2}}{\sqrt{2\pi t}}%
,\;{\rm orthogonal}
\end{equation}
\begin{equation}
\label{wf8}f_{X\rightarrow \infty }\left( t\right) =e^{-t},\;{\rm unitary}
\end{equation}
Eqs. (\ref{wf7}, \ref{wf8}) for the distribution functions of chaotic
electron waves in the limiting cases of the orthogonal and unitary ensembles
agree with the corresponding results obtained with the help of eigenvectors
projections in the RMT \cite{JBS,Bro} and confirm the result for the unitary
ensemble derived from the analysis of the participation ratios in Refs. \cite
{EP,Pri}. Recently, an independent calculation of the function $f\left(
t\right) $ starting from a model with random matrices and using a different
method was performed in Ref. \cite{IiS}. The final representation of the
function $f\left( t\right) $ in Ref. \cite{IiS} differs from Eqs. (\ref{wf6}%
, \ref{wf6a}). Although the asymptotics Eqs. (\ref{wf7}, \ref{wf8}) are
reproduced in Ref. \cite{IiS} it is not yet clear whether Eqs.(\ref{wf6},
\ref{wf6a}) coincide with the corresponding equations of Ref. \cite{IiS} or
not.

The crossover between the orthogonal and unitary ensembles occurs at
relatively small magnetic fluxes through the area of the sample, $\phi \sim
\phi _0\left( \Delta /E_c\right) ^{1/2}.$ The probability of the zero
amplitude $|\psi |$ of the wave function decays fast when a weak magnetic
field is applied to the system
\begin{equation}
\label{wf9}f_X\left( 0\right) =2\sqrt{\pi }/\left( 3X\right) ,\text{ }\;X\ll
1
\end{equation}
The exponentially rare events of finding a pronounced splash of the single
state electron density above the average level of $V^{-1}$ can be found from
Eq. (\ref{wf6}) using the saddle-point method of evaluating the integral
over $x.$ With an exponential accuracy it can be approximated as
\begin{equation}
\label{wf10}f_X\left( t\gg 1\right) \propto \exp \left\{ -\frac t2\left[ 1+%
\frac{X^2}t\left( \sqrt{1+\frac{2t}{X^2}}-1\right) \right] \right\}
\end{equation}
In the limits $X\ll 1$ and $X\gg 1$ Eq. (\ref{wf10}) gives the correct
exponents, Eqs. (\ref{wf7}, \ref{wf8}). The distribution function $f\left(
t\right) $ derived here used together with Eq. (\ref{st10}) enables us to
derive the resonance distribution function. The calculation is
straightforward but has not been done yet.

\section{Conclusion}

The supersymmetry technique developed originally to study disordered metals
turns out to be a powerful tool for studying problems of quantum chaos and
mesoscopic physics. The zero- dimensional version of the $\sigma $-model is
equivalent to the Gaussian ensembles of the random matrix theory. The
applicability of the latter for studying the quantum chaos is the basis of
using the supersymmetry method for calculation of different correlation
functions of the chaotic systems. The $\sigma $-model can be derived from
the first principles from model with the random potential, thus proving the
applicability of the RMT to investigation of small disordered metal
particles. However, calculations with the supersymmetry method are more
straightforward and one has to calculate in this method integrals over
several variables instead of integrals over arbitrarily large number of
variables as in the RMT. In the last Sections calculation of some
distribution functions have been presented. With the exception of the
function $f\left( t\right) $ in the limiting cases of the orthogonal and
unitary ensembles the above distribution functions have not been known in
the RMT and the corresponding problems have not even been formulated.

The supersymmetry method becomes even more efficient when problems
corresponding to higher dimensions are considered. For example, the problem
of ''kicked rotator'' and other problems where the the phenomenon of dynamic
localization shows up can be described by the random band matrices and,
which is equivalent, by the one dimensional $\sigma $-model. Some models
with ''sparse random matrices'' can be mapped onto the $\sigma $-model on
the Bethe lattice \cite{FM}. The $\sigma $-model on the Bethe lattice
possesses the metal-insulator transition and has been solved exactly some
time ago \cite{BL,Zir2}. All these examples show that the number of
applications of the supersymmetry technique is growing and I can only
express the hope that many new interesting problems will be solved by this
method.

\begin{references}
\bibitem{Gian}  M.-Giannoni, A. Voros, and J. Zinn-Justin (editors), {\it %
Chaos and Quantum Physics (}North-Holland, New York, 1991)

\bibitem{Gut}  M.C.Gutzwiller, {\it Chaos in Classical and Quantum Mechanics
}(Springer-Verlag, New York, 1991)

\bibitem{Wig}  E.P. Wigner, {\it Ann. Math.}, {\bf 53,} 36, (1951)

\bibitem{Dy}  F.J. Dyson, {\it J. Math. Phys.,}{\bf 3}, 140, 157, 166,
(1962) {\it \ }

\bibitem{Meh}  M.L. Mehta, {\it Random Matrices (}Academic Press Inc., New
York, 1990)

\bibitem{Ber}  M.V. Berry, {\it Proc. R. Soc. Lond. A }{\bf 400, }229, {\bf (%
}1985)

\bibitem{Smy}  U. Smilansky in {\it Chaos and Quantum Physics, }see ref.
\cite{Gian}

\bibitem{Arg}  N. Argaman, J. Imry, U. Smilansky, {\it Phys. Rev. B, }{\bf %
47, }4440, (1993)

\bibitem{Jal}  R.A. Jalabert, H.U. Baranger, and A.D. Stone, {\it Phys. Rev.
Lett.} {\bf 65, } 2442, (1990); H.U. Baranger, R.A. Jalabert, and A.D.
Stone, {\it Chaos, {\bf 3, }} 665, (1993){\it \ }

\bibitem{Bee}  C.W.J. Beenakker and H. van Houten in {\it Solid State
Physics, }edited by H. Ehrenreich and D. Turnbull (Academic, New York, 1991)

\bibitem{Alt}  B.L. Altshuler, P.A. Lee, and R.A. Webb (editors), {\it %
Mesoscopic Phenomena in Solids }(North-Holland, New York, 1991)

\bibitem{Efe}  K.B. Efetov, {\it Adv. in Physics,} {\bf 32,} 53, (1983)

\bibitem{Ver}  J.J.M. Verbaarschot, H.A. Weidenm\"uller and M.R. Zirnbauer,
{\it Phys. Reports, }{\bf 129, } 367, (1985)

\bibitem{Altl}  A. Altland, S. Iida and K.B. Efetov, {\it J. Phys. A }{\bf %
26, } 3545, (1993)

\bibitem{Izr}  F.M. Izrailev, {\it Phys. Rep. }{\bf 129, } 2999, (1990)

\bibitem{Cas}  G. Casati, B.V. Chirikov, I. Guarnery, D.L. Shepelyansky,
{\it Phys. Rep. }{\bf 154,} 77, (1987)

\bibitem{CMI}  G. Casati, L. Molinary, F.M. Izrailev, {\it Phys. Rev. Lett. }%
{\bf 64, }16, (1990); M. Feingold, D.M. Leitner, M.Wilkinson, {\it Phys.
Rev. Lett. }{\bf 66,} 986, (1991)

\bibitem{Fyo}  Y.V. Fyodorov, A.V. Mirlin, {\it Phys. Rev. Lett. }{\bf 67,}
2405, (1991); {\it Phys. Rev. Lett. }{\bf 69,} 1093, (1992)

\bibitem{Sim}  B.D. Simons, B.L. Altshuler, {\it Phys. Rev. Lett. }{\bf 70, }%
4063 (1993); {\it Phys. Rev. B }{\bf 48,} 5422, (1993)

\bibitem{Lee}  B.D. Simons, P.A. Lee, B.L. Altshuler, {\it Nucl. Phys. B, }%
{\bf 409, }487, (1993)

\bibitem{EP}  K.B. Efetov, V.N. Prigodin, {\it Phys. Rev. Lett. }{\bf 70, }%
1315, (1993); {\it Mod. Phys. Lett. }{\bf 15,} 981, (1993)

\bibitem{Hal}  W.P. Halperin, {\it Rev. Mod. Phys. }{\bf 58,} 533, (1986)

\bibitem{Brom}  H.B. Brom, D. van der Putten, L.J. de Jong, in {\it Physics
and Cemistry of Metal Cluster Compounds. Model Systems for small Metal
Particles (}Riedel, Dordrecht, to be published in 1994)

\bibitem{Bee2}  C.W.J. Beenakker, preprint (1994)

\bibitem{Yee}  P. Yee, W.D. Knight, {\it Phys. Rev. }{\bf 11,} 3261, (1975)

\bibitem{Brom2}  H.B. Brom, F.C. Fritschij, D. van Putten, F.A. Hanneman,
L.J. de Jong, and G. Schmid, preprint, (1994)

\bibitem{Kirk}  W.P. Kirk and M.A. Read (editors), {\it Nanostructures and
Mesoscopic Physics, }(Academic Press, San Diego, 1992){\it \ }

\bibitem{Kas}  M.A. Kastner, {\it Rev. Mod. Phys. }{\bf 64, }849, (1992)

\bibitem{Mar}  C.M. Marcus, A.J. Rimberg, R.M. Westervelt, P.F. Hopkins and
A.G. Gossard, {\it Phys. Rev. Lett. }{\bf 69, }506, (1992); C.M. Marcus,
R.M. Westervelt, P.F. Hopkins and A.C. Gossard, {\it Phys. Rev. B, }{\bf 48,
}2460 (1993); C.M. Marcus, R.M. Westervelt, P.F. Hopkins and A.G. Gossard,
{\it Chaos, }{\bf 3,} 643, (1993)

\bibitem{Pri}  V.N. Prigodin, K.B. Efetov, and S. Iida, {\it Phys. Rev.
Lett. }{\bf 71, }1230, (1993)

\bibitem{Iid}  S. Iida, H.A. Weidenm\"uller, and J.A. Zuk, {\it Annals of
Physics, }{\bf 200, }219, (1991)

\bibitem{Zir}  M.R. Zirnbauer, {\it Nucl. Phys.A }{\bf 560,}95, (1993)

\bibitem{ALS}  B.L. Altshuler, {\it Pis}`ma Zh. Eksp. Teor. Fiz. {\bf 41,}
530, (1985) ({\it JETP Lett. }{\bf 41,} 648, (1985)); P.A. Lee and A.D.
Stone , {\it Phys. Rev. Lett, }{\bf 55, }1622, (1985)

\bibitem{JBS}  R.A. Jalabert, A.D. Stone, and Y. Alhassid, {\it Phys. Rev.
Lett. }{\bf 68,} 3468, (1992)

\bibitem{Bro}  T.A. Brody, J. Flores, J.B. French, P.A. Mello, A. Pandy, and
S.S.M. Wong, {\it Rev. Mod. Phys. }{\bf 53,} 385, (1981)

\bibitem{Fal}  V.I. Falko, K.B. Efetov, {\it Phys. Rev. B , }to be published
(1994)

\bibitem{IiS}  H.J. Sommers, S. Iida, {\it Phys. Rev. E, }{\bf 49, }2513,
(1994)

\bibitem{FM}  Y.V. Fyodorov, A.D. Mirlin,{\it \ Phys. Rev. Lett. }{\bf 67,}
2049, (1991)
8
\bibitem{BL}  K.B. Efetov, {\it Zh. Eksp. Teor. Fiz. }{\bf 88, }1032, (1984)
({\it Sov. Phys. JETP, }{\bf 61, }606, (1984)); {\it Zh. Eksp. Teor. Fiz. }%
{\bf 92,} 638, 1987 ({\it Sov. Phys. JETP }{\bf 65,} 360, (1987)); {\it Zh.
Eksp. Teor. Fiz. }{\bf 93, }1125, (1987) ({\it Sov. Phys. JETP }{\bf 66,}
634, 1987);

\bibitem{Zir2}  M.R. Zirnbauer, {\it Nucl. Phys. B, }{\bf 265, }375, (1986);
{\it Phys. Rev.} {\bf 34, }6394, (1986)
\end{thebibliography}

\end{document}